# ON THE EFFICIENCY OF TRANSMISSION STRATEGIES FOR BROADCAST CHANNELS USING FINITE SIZE CONSTELLATIONS

*Zeina Mheich, Florence Alberge and Pierre Duhamel*

LSS - Supelec - Univ Paris-Sud
3 rue Joliot-Curie, 91192 Gif-sur-Yvette cedex, France
E-mail: {zeina.mheich, alberge, pierre.duhamel}@lss.supelec.fr

**ABSTRACT**

In this paper, achievable rates regions are derived for power constrained Gaussian broadcast channel of two users using finite dimension constellations. Various transmission strategies are studied, namely superposition coding (SC) and superposition modulation (SM) and compared to standard schemes such as time sharing (TS). The maximal achievable rates regions for SM and SC strategies are obtained by optimizing over both the joint probability distribution and over the positions of constellation symbols. The improvement in achievable rates for each scheme of increasing complexity is evaluated in terms of SNR savings for a given target achievable rate or/and percentage of gain in achievable rates for one user with reference to a classical scenario.

*Index Terms*— AWGN broadcast channels, achievable rates region, superposition coding, signal constellations, shaping gains.

## 1. INTRODUCTION

Hierarchical modulation or layered modulation is used in various standards such as DVB-T/H/SH in order to help various users of different channel conditions to obtain a given quality of service. It enables the transmission of multiple services simultaneously on a single frequency channel with different transmission qualities.

The first question that arises here is about the optimality of the achievable rates when using hierarchical modulation as a transmission strategy. It is known since Cover [1], that the theoretical limit of the capacity region for two users Gaussian broadcast channel is achieved using superposition coding with infinite Gaussian input alphabets. Clearly, this is not the case in hierarchical modulation for which the restriction imposed in using finite signaling constellation of equiprobable symbols reduces the achievable rates and leads to a gap with the capacity region. This gap can be reduced using constellation shaping. For two-users Gaussian broadcast channels, the idea of constellation shaping is to maximize the achievable rates region by optimizing over both the joint pdf and the positions of constellation symbols. In fact, most results for constellation shaping with finite signal constellations considered only point to point communication systems [2]-[6].

For broadcast channels, the achievable rates region for two-user AWGN broadcast channels with finite input alphabets are derived in [7] when superposition of modulated signal points is used assuming a uniform distribution over the finite input set. In [8] and [9], the problem of constellation shaping for Gaussian broadcast channels is introduced when the transmitted signal is modulated using a 4-PAM constellation. This problem was investigated only for a particular situation where each user can receive at most two symbols. This corresponds to the case with the least complexity, since the receiver can identify his information using the labeling. Noticeable shaping gain of up to 2 dB is obtained on the SNR of one user for a fixed SNR of the other user.

Constellation shaping for broadcast channels using simple transmission scheme is shown to be useful in some situations depending on users channel conditions and for low order constellation 4-PAM [8][9]. In this paper, we extend this work to study achievable rates using {4,8,16}-PAM constellations as well as superposition coding (SC, i.e. the most general case, to be defined below). Hence, achievable rates regions are given here for superposition coding, superposition modulation and time sharing. Constellation shaping is used to derive maximal achievable rates for SM and SC strategies. The loss experienced by using simple schemes is evaluated and then the situations in which more decoding complexity allows significant improvements are identified.

## 2. TWO-USER AWGN BROADCAST CHANNEL

A two-user broadcast channel (BC) consists of an input alphabet $\mathcal{X}$, two outputs alphabets $\mathcal{Y}_1$ (user 1), $\mathcal{Y}_2$ (user 2) and a conditional pdf $P_{Y_1Y_2|X}$ on $\mathcal{Y}_1 \times \mathcal{Y}_2$. Let $X$, $Y_1$ and $Y_2$ be random variables representing the input and outputs of the BC. A BC is said to be physically degraded if $P_{Y_1Y_2|X}(y_1,y_2|x) = P_{Y_1|X}(y_1|x) \cdot P_{Y_2|Y_1}(y_2|y_1)$ (i.e. $X \rightarrow Y_1 \rightarrow Y_2$ form a Markov chain). A BC is said to be stochastically degraded or degraded if there exists a random variable $\tilde{Y}_1$ which has the same conditional pdf as $Y_1$ given $X$ such that $X \rightarrow \tilde{Y}_1 \rightarrow Y_2$ form a Markov chain.

The private-message capacity region of the degraded broadcast channel $X \to Y_1 \to Y_2$ is the convex hull of the closure of rate pairs $(R_1, R_2)$ satisfying:

$$\begin{array}{rcl} R_1 & \leq & I(X;Y_1|U) \\ R_2 & \leq & I(U;Y_2) \end{array} \quad (1)$$

for some joint distribution $P_{UXY_1Y_2} = P_{UX} \cdot P_{Y_1|X} \cdot P_{Y_2|X}$ on $\{\mathcal{U} \times \mathcal{X} \times \mathcal{Y}_1 \times \mathcal{Y}_2\}$. $P_{Y_1|X}$ and $P_{Y_2|X}$ are the conditional pdf that depend on the nature of the channel. $P_{UX}$ is the joint probability distribution of $U$ and $X$, where the auxiliary random variable $U$ has cardinality bounded by $|\mathcal{U}| \leq \min\{|\mathcal{X}|, |\mathcal{Y}_1|, |\mathcal{Y}_2|\}$. The capacity region is achieved using superposition coding described next.

We consider in this work the Gaussian BC with two users. Without loss of generality, we assume in this paper that $Y_1$ is less noisy than $Y_2$. It can easily be shown that scalar Gaussian broadcast channels are equivalent to a degraded channel [10],

$$\begin{array}{l} Y_1 = X + Z_1 \\ Y_2 = X + Z_2 = Y_1 + Z_2' \end{array}$$

where $Z_1 \sim \mathcal{N}(0, \sigma_1^2)$, $Z_2 \sim \mathcal{N}(0, \sigma_2^2)$, $Z_2' \sim \mathcal{N}(0, \sigma_2^2 - \sigma_1^2)$, $\sigma_i^2$ is the variance of the noise $Z_i$ and $Z_1, Z_2'$ are independent. Thus Gaussian BC is stochastic degraded. We assume an average power constraint on the transmitted power $P$ defined as $\mathbb{E}[X^2] \leq P$. The received signal to noise ratio for each user $i$ is $SNR_i = \frac{P}{\sigma_i^2}$.

## 3. BROADCAST TRANSMISSION STRATEGIES

In this section, we introduce briefly various transmission strategies for broadcast systems:

### 3.1. Time sharing (TS)

In time sharing scheme, a percentage of time is used to send data for one user and the rest of the time is used to send for the other user. In this case, the rate pairs can be achieved by strategies used for point to point channel and sharing the time between users.

### 3.2. Superposition coding (SC)

The basics of superposition coding are briefly recalled below; a detailed description is given in [10]. In superposition coding, some auxiliary random variable $U$ serves as a cloud center distinguishable by both receivers [1][10]. Each cloud consists of $2^{nR_1}$ codewords $X^n$ of length $n$ distinguishable by receiver 1. Receiver 2 only sees the clouds while receiver 1 can see the individuals codewords within the clouds. To generate the codebook, first choose the joint distribution $P_{UX} = P_U P_{X|U}$. Then, from the message $w_2 \in \{1, ..., 2^{nR_2}\}$, generate a codeword $u^n(w_2)$ according to $P_U$ which is distinguishable by both receivers. For each codeword $u^n(w_2)$ and $w_1 \in \{1, ..., 2^{nR_1}\}$, generate codeword $x^n(w_1, w_2)$ according to $P_{X|U}$ where the extra information in $x^n$ is viewed as noise by receiver 2. The encoding is based on transmitting $x^n(w_1, w_2)$ to send the message pair $(w_1, w_2)$ and the decoding is based on joint typicality. Since the auxiliary random variable $U$ has cardinality bounded by $|\mathcal{U}| \leq \min\{|\mathcal{X}|, |\mathcal{Y}_1|, |\mathcal{Y}_2|\}$, we use the name general superposition coding or superposition coding simply to describe the case where $|\mathcal{U}| = \min\{|\mathcal{X}|, |\mathcal{Y}_1|, |\mathcal{Y}_2|\}$. Cover [1] showed that in the case of binary symmetric BC and AWGN BC, superposition coding expands the rate region beyond that achievable with time sharing. For SC and with $M$-PAM modulation of $M$ points, $P_{UX}$ is a $M \times M$ matrix with taps $p_{i,j}$. Up to our knowledge, no study have been presented about the derivation of achievable rates for the general superposition coding using finite order constellations.

### 3.3. Superposition modulation (SM)

This is a special case of superposition coding scheme. In this case, $2^{nR_2}$ independent codewords $u^n(w_2)$ are generated according to $P_U$ and for each of these codewords, $2^{nR_1}$ satellite codewords $v^n$ are generated and added to form codewords $x^n(w_1, w_2) = u^n + v^n$ according to $P_{X|U}$. Thus, the fine information $v^n$ is superimposed on the coarse information $u^n$. The capacity region of Gaussian broadcast channel is achieved using this coding scheme and successive cancellation decoding where $U$ and $V$ are independent random variables following normal distributions. However, we don't assume here that $U$ and $V$ should be independent in superposition modulation. In SM, $P_{UX}$ takes a specific expression which give the corresponding labeling of the $M$-PAM constellation for a fixed labeling for $\mathcal{X}_1$ and $\mathcal{X}_2$, the alphabets for users 1 and 2 respectively [8][9]. We consider, as an example, a 8-PAM modulation. In that case, the transmitted signal is the sum of the two users signals and is given by $x = x^{(1)} + x^{(2)}$ where $x^{(1)} \in \mathcal{X}_1$ and $x^{(2)} \in \mathcal{X}_2$ with $M_1 = |\mathcal{X}_1|$, $M_2 = |\mathcal{X}_2|$ and $M_1 M_2 = 8$. This leads to $U \equiv X_2$ and $V \equiv X_1$ for superposition modulation, where $X_i$ takes values from $\mathcal{X}_i$, since user 2 can distinguish only $U$. Two configurations are possible either $M_2 = 4$ ($\mathcal{X}_1$ is a BPSK and $\mathcal{X}_2$ is a 4-PAM) or $M_2 = 2$ ($\mathcal{X}_1$ is a 4-PAM and $\mathcal{X}_2$ is a BPSK). In these two cases, $P_{UX}$ is a sparse matrix of size $M_2 \times M$ with expressions

$$P_{UX} = \begin{bmatrix} p_{00} & p_{01} & 0 & 0 & 0 & 0 & 0 & 0 \\ 0 & 0 & p_{12} & p_{13} & 0 & 0 & 0 & 0 \\ 0 & 0 & 0 & 0 & p_{24} & p_{25} & 0 & 0 \\ 0 & 0 & 0 & 0 & 0 & 0 & p_{36} & p_{37} \end{bmatrix}$$

$$if\ M_1 = 2, M_2 = 4 \quad (2)$$

$$P_{UX} = \begin{bmatrix} p_{00} & p_{01} & p_{02} & p_{03} & 0 & 0 & 0 & 0 \\ 0 & 0 & 0 & 0 & p_{14} & p_{15} & p_{16} & p_{17} \end{bmatrix}$$

$$if\ M_1 = 4, M_2 = 2 \quad (3)$$

where $P_{UX}[i,j] = p_{i-1,j-1} = \Pr\{U = u_{i-1}, X = x_{j-1}\}$. In both cases, the number of taps to be computed is 8.

## 4. ACHIEVABLE RATES REGIONS COMPUTATION

Consider a two user memoryless AWGN broadcast channel ($SNR_1 > SNR_2$) with signal power constraint $P$. The channel input lies in a finite set $\mathcal{X} = \{x_0, ..., x_{M-1}\} \subset \mathbb{R}$ represented by a $M$-PAM constellation. Assume a symmetric input signal constellation with respect to the origin. Since $\mathcal{U}$ has cardinality bounded by $|\mathcal{U}| \leq \min\{|\mathcal{X}|, |\mathcal{Y}_1|, |\mathcal{Y}_2|\}$ and the output alphabet cardinality for an AWGN channel is infinite, we have $|\mathcal{U}| \leq |\mathcal{X}|$. Thus $|\mathcal{U}| \leq M$. To determine the maximal achievable rates region using SC, we consider the case $|\mathcal{U}| = M$. However for SM, we take into account the specificities on the joint pdf $P_{UX}$ given in section 3.3. The strategies using SC or SM under consideration are given in table 1. To determine the maximal achievable rates region for each strategy in table 1, we solve the following optimization problem for $\theta \in [0, 1]$ where $\theta$ is the weight given to user 1:

$$\max_{P_{UX}, \mathcal{X}} f(P_{UX}, \mathcal{X}) = \theta \cdot I(X; Y_1|U) + (1-\theta) \cdot I(U; Y_2) \quad (4)$$

subject to a) the power constraint $\sum_{i,j} p_{ij} \cdot x_j^2 \leq P$ and b) the constraint on the joint pdf $P_{UX}$ given in table 1 for each strategy, where $p_{ij} = \Pr\{U = u_i, X = x_i\}$. To solve this optimization problem, we form its Lagrangian $L(P_{UX}, \mathcal{X}, s) = f(P_{UX}, \mathcal{X}) + s \cdot (P - \sum_{i,j} p_{ij} \cdot x_j^2)$. For a given value of $s$, the optimization of the Lagrangian is solved with respect to $P_{UX}$ (verifying the constraint b)) and $\mathcal{X}$ alternately until convergence. We repeat this process until finding the optimal value of $s$ for which the power constraint is fulfilled.

For the TS using standard $M$-PAM, the couple of achievable rates is given by ($R_1 = \alpha \overline{R_1}$, $R_2 = (1-\alpha) \overline{R_2}$), where $\overline{R_1}$ and $\overline{R_2}$ are achievable rates for point to point channel using standard $M$-PAM constellation at $SNR_1$ and $SNR_2$ respectively. A standard $M$-PAM constellation is defined as a constellation with $M$ equally probable real symbols belonging to $\mathcal{X} = \{M - 1 - 2 \cdot (i-1), \text{for } i = 1, ..., M\}$. Varying $\alpha$ from 0 to 1 yields achievable rates region.

In this work, we are interested in the case where the message $w_2$ is a common message to both receivers. However $w_1$ is a private message to user 1. Consequently, user 1 achieves a rate $R_1 + R_2$ while user 2 achieves a rate $R_2$. So it is sufficient to solve problem (4) for $\theta \in [0, \frac{1}{2}]$.

## 5. SIMULATION RESULTS AND DISCUSSION

In [8] and [9], constellation shaping for Gaussian BC is studied only for SM and using a low order constellation 4-PAM. In this section, we extend this work to analyze the achievable rates for an $M$-PAM constellation with $M \in \{4, 8, 16\}$ and for other transmission strategies including the general case of superposition coding. Precisely, all the schemes in table 1 will be considered as well as TS.

| Tx | Variables and constraints in (4) | Designation |
|---|---|---|
| SC | $\mathcal{X}$ and $P_{UX}$ s.t. $\sum_{i,j} p_{i,j} = 1$ | $SC_{\mathcal{X}, P_{UX}, P_X}$ |
| SM | $\mathcal{X}$ and $P_{UX}$ s.t. $\sum_{i,j} p_{i,j} = 1$ | $SM_{\mathcal{X}, P_{UX}, P_X}$ |
| SM | $\mathcal{X}$ and uniform $P_{UX}$ | $SM_{\mathcal{X}, \overline{P_{UX}}, \overline{P_X}}$ |

**Table 1**. Strategies under consideration

Achievable rates region curves are given in Fig.1 for $M = 4, 8, 16$. For each value of $M$, the display of the results is limited to one couple of $SNR$. In complement with the achievable rates region curves, comparisons are also conducted in terms of SNR savings for target achievable rates (Maximum Shaping Gain) and in terms of Maximum Percentage of Gain for user 1. These two quantities are defined below. Let consider two transmission strategies ($A$ and $B$):

**Definition 1.** *The pair of rates $(R_1 + R_2, R_2)$ is achieved for $(SNR_1, SNR_2)$ with A and for $(SNR_1 + \Delta SNR, SNR_2 + \Delta SNR)$ with B. The shaping gain (with A compared to B) is $\Delta SNR$. The maximum shaping gain is defined as:*

$$MG_{SNR_{dB}}(A|B) = max_{R_2} \Delta SNR$$

**Definition 2.** *For a given pair of SNR $(SNR_1, SNR_2)$ and a fixed value of $R_2$, the achievable pair of rates is $(R_1^A + R_2, R_2)$ resp. $(R_1^B + R_2, R_2)$ with A resp. B. The gain on the achievable rate for user 1 is given by*

$$G_{R_1}(A|B) = \frac{(R_1^A + R_2) - (R_1^B + R_2)}{R_1^B + R_2} \cdot 100 \, (\%)$$

*The maximum gain on the achievable rate for user 1 (with A compared to B) is given by $MG_{R_1}(A|B) = max_{R_2} G_{R_1}(A, B)$*

### 5.1. Superposition modulation

In this section, the two configurations of superposition modulation are compared. Figures of achievable rates regions show that an improvement can be obtained with $SM_{\mathcal{X}, P_{UX}, P_X}$ (full optimization) compared to $SM_{\mathcal{X}, \overline{P_{UX}}, \overline{P_X}}$ and depending on $\delta_{SNR} = SNR_1 - SNR_2$. To quantify this improvement, the maximum gain in achievable rate ($MG_{R_1}$) and the maximum SNR savings ($MG_{SNR_{dB}}$) are given in table 2. We observe the following. A slightly gain in terms of achievable rates can be translated into a noticeable gain in terms of $SNR$ saving. The maximum shaping gain increases with the constellation size ($M$). Thus, constellation shaping for SM strategy seems more useful for high values of $M$. The analysis of the optimal matrix $P_{UX}$ (results not reported) leads to the conclusion that $X_1$ and $X_2$ are not independent in general when using finite-size constellations. We observe also that the maximum shaping gain for $SM_{\mathcal{X}, P_{UX}, P_X}$ versus $SM_{\mathcal{X}, \overline{P_{UX}}, \overline{P_X}}$

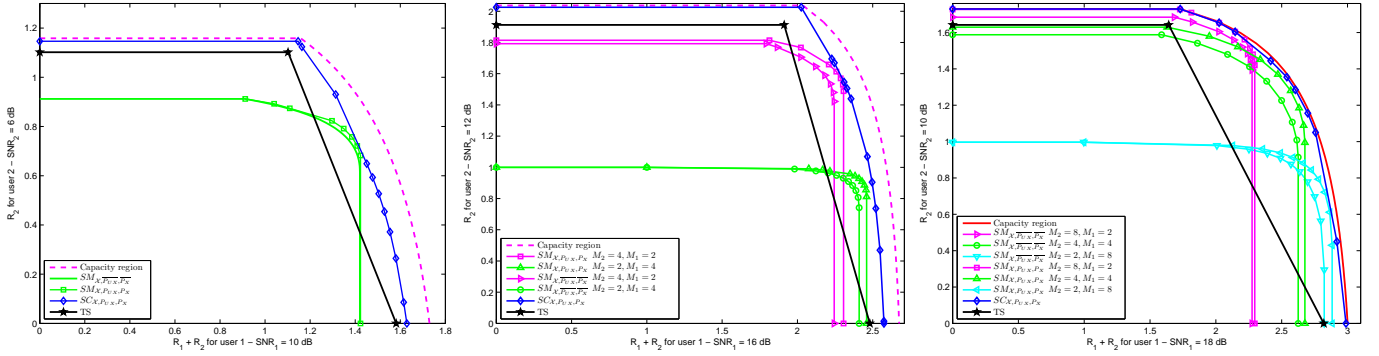

**Fig. 1**. Achievable rate regions with: (a) $M = 4$ - (b) $M = 8$ - (c) $M = 16$

| $M$ | $SNR_1$ | $SNR_2$ | $MG_{SNR_{dB}}$(A|B) | $MG_{R_1}$(A|B) |
|---|---|---|---|---|
| 4 | 10 | 8 | 0.39 | 7.46% |
|   |    | 6 | 0.17 | 3.51% |
|   |    | 4 | 0.05 | 1.77% |
|   |    | 2 | 0.01 | 0.38% |
| 8 | 16 | 14 | 0.71 | 20.17% |
|   |    | 12 | 0.57 | 13.21% |
|   |    | 10 | 0.41 | 13.07% |
|   |    | 8  | 0.33 | 18.93% |
| 16 | 18 | 16 | 1.05 | 10.67% |
|    |    | 14 | 0.87 | 11.54% |
|    |    | 12 | 0.64 | 12.08% |
|    |    | 10 | 0.49 | 19.53% |

**Table 2**. Comparison of $SM_{\mathcal{X},P_{UX},P_X}$ (A) and $SM_{\mathcal{X},\overline{P_{UX},P_X}}$) (B) with respect to $MG_{SNR_{dB}}$ and $MG_{R_1}$

increases when $\delta_{SNR}$ decreases, independently of $M$. In particular full optimization (vs optimization of the symbol position) is insignificant for large $SNR$ gap in SM strategy.

### 5.2. Time-sharing or superposition modulation?

This section compares two strategies (TS and SM) classically considered in broadcast systems. In Fig.1.a ($M = 4$), we observe that the achievable rates region can be split into 2 parts. Indeed, for small and large values of $R_2$, TS is better than SM. On the contrary, SM is better than TS for middle-range values of $R_2$. Under a given rate requirement for one user, we can thus determine the best transmission strategy. We can also observe that the region in which SM is better than TS becomes small for low values of $SNR_2$. With $M = 8$ (Fig.1.b), the area in which SM is better than TS increases (compared to $M = 4$) by considering the union of the two possible configurations for SM: $M_1 = 2$, $M_2 = 4$ (case 1) and $M_1 = 4$, $M_2 = 2$ (case 2). This is particularly true when $\delta_{SNR}$ increases. We also observe that TS can achieve higher rates than SM (case 1) for good $SNR_2$ values. Indeed, the maximum rate of user 2 with SM is the maximum individual rate for a 4-PAM constellation whereas it is the individual user rate achieved using standard 8-PAM in the TS case. For low $SNR_2$ values optimized 4-PAM may achieve higher rate than standard 8-PAM thus SM becomes better in this interval. For a 16-PAM constellation (Fig.1.c), SM is always better than TS for the studied couples of ($SNR_1$, $SNR_2$). Table 3 show maximum percentage of improvement in achievable rate of user 1 by TS when using $SM_{\mathcal{X},P_{UX},P_X}$ strategy in the interval where $SM_{\mathcal{X},P_{UX},P_X}$ is better than TS. Clearly, the maximum percentage of improvement increases when $\delta_{SNR}$ increases and an important gain is obtained for high values of $\delta_{SNR}$ as in the case of $SNR_1 = \delta_{SNR} = 10dB$ for a 4-PAM where the percentage of gain on achievable rate of user 1 varies between 0 and $40.7\%$. For a 8-PAM constellation, the percentage of gain on achievable rate of user 1 varies between 0 and $30.21\%$ when $SNR_1 = 16$ $dB$ and $\delta_{SNR} = 8$ $dB$. For a 16-PAM, percentages of improvements can be up to $35.08\%$ when $SNR_1 = 18dB$ and $\delta_{SNR} = 8dB$. We can conclude that SM is a better option than TS especially for large $\delta_{SNR}$ values. TS is optimal in the region where we want to maximize the rate of user 2 for good values of $SNR_2$ because the single user rate achieved by TS is the rate achieved using standard $M$-PAM constellation (the constellation is split between users with SM). Thus, SM seems more gainful than TS when we want to serve many users.

### 5.3. Superposition coding

For the three constellations under consideration ($M = 4, 8, 16$), the maximal achievable rates region obtained by the optimal general case of superposition coding when we consider the general form of $P_{UX}$ (SC) achieves an important region of rates couples ($R_1 + R_2$, $R_2$) that cannot be achieved neither by TS nor by SM. Even when we optimize SM ($SM_{\mathcal{X},P_{UX},P_X}$) we are far from maximal achievable rates region. Sometimes the maximal achievable rates region curve is very close or even coincides with the $SM_{\mathcal{X},P_{UX},P_X}$ achiev-able rates region in a couple of rates. This is the case when $SM_{\mathcal{X},P_{UX},P_X}$ is the optimal superposition coding in terms of achievable rates. We are interested now in the numerical evaluation of the gain in rate of user 1 ($R_1 + R_2$) when we use $SC_{\mathcal{X},P_{UX},P_X}$ compared to the best strategy between TS and SM. This gain ($MG_{R_1}(SC_{\mathcal{X},P_{UX},P_X}|TS \bigcup SM_{\mathcal{X},P_{UX},P_X})$) calculated in % is the distance between the limit of the maximal achievable rates region and the limit of closure of achievable rates region of TS and $SM_{\mathcal{X},P_{UX},P_X}$. The results are reported in table 3. We observe that the part of the maximal achievable rates region which is unachievable by TS and SM, is bigger when $M$ is small because we observe that for the case of 4-PAM we have one configuration for SM. However, we have two configurations of SM for 8-PAM constellation and three configurations for 16-PAM constellation. Thus when $M$ increases, the union of achievable rates for all SM cases tends to the sets of achievable rates by the general superposition coding. Asymptotically, we know that when $M \to \infty$, $SM_{\mathcal{X},P_{UX},P_X}$ is the optimal superposition coding scheme because it allows to achieve the capacity region for two-user AWGN BC using Gaussian alphabet for each user. Thus the maximum gain in user 1 rate decreases when constellation order $M$ increases. We observe also that the gain in achievable rates is high for high values of $\delta_{SNR}$.

| $M$ | $SNR_1$ | $SNR_2$ | $MG_{R_1}(A|B)$ | $MG_{R_1}(A|C)$ |
|---|---|---|---|---|
| 4 | 10 | 8 | 6.13% | 6.72% |
| | | 6 | 11.14% | 11.65% |
| | | 4 | 18.50% | 16.69% |
| | | 2 | 28.43% | 18.9% |
| | | 0 | 40.70% | 23.54% |
| 8 | 16 | 14 | 7.80% | 7.89% |
| | | 12 | 13.60% | 11.43% |
| | | 10 | 21.15% | 14.96% |
| | | 8 | 30.21% | 14.71% |
| 16 | 18 | 16 | 10.36% | 2.96% |
| | | 14 | 16.42% | 2.94% |
| | | 12 | 24.68% | 5.29% |
| | | 10 | 35.08% | 4.80% |

**Table 3**. Comparison of $SM_{\mathcal{X},P_{UX},P_X}$ (A) vs $TS$ (B). Comparison of $SC_{\mathcal{X},P_{UX},P_X}$ (A) vs $TS \bigcup SM_{\mathcal{X},P_{UX},P_X}$ (C).

In general we conclude that fixing constellations of users (i.e. assigning labels to the constellation so that we distinguish between the bits intended for each user) is not optimal for coding and may result in important loss in terms of rates for systems using finite-size constellations especially for low-order constellations. Which is better, is to determine the optimal alphabet of the auxiliary alphabet $U$ which is not necessarily a constellation and then to generate the codewords $x^n$ which are not necessarily the sum of two codewords (this scheme is explained in paragraph 3.2).

## 6. CONCLUSION

In this paper, we derived achievable rates region for power constrained AWGN BC of two users using $M$-PAM constellations and for various broadcast transmission strategies. The maximal achievable rates region for SC and SM are obtained using constellation shaping. For SM, results showed that constellation shaping seems more useful for high values of $M$. Moreover, the gain in using a complex case of SM increases when the SNR gap between users decreases. We observe also that SM outperforms time sharing in a large part of the achievable rates region. Finally, it is shown that using the general case of superposition coding can bring important gains comparing to classical schemes especially for small values of $M$.